\documentclass[twocolumn,preprintnumbers,amsmath,amssymb,superscriptaddress]{revtex4}
\usepackage{amssymb}
\usepackage{graphicx,color}
\usepackage{bm}

\begin{document}

\title{Transition balance in QCD nucleation}

\author{Tianzhe Zhou}
\affiliation{Department of Physics, Tsinghua University, Beijing 100084, China}

\author{Qiuze Sun}
\affiliation{Department of Physics, Tsinghua University, Beijing 100084, China}

\author{Jin Hu}
\affiliation{Department of Physics, Fuzhou University, Fujian 350116, China}

\author{Carsten Greiner}
\affiliation{Institut f$\ddot{u}$r Theoretische Physik, Johann Wolfgang Goethe-Universit$\ddot{a}$t Frankfurt, Max-von-Laue-Strasse 1, 60438 Frankfurt am Main, Germany}

\author{Zhe Xu}
\affiliation{Department of Physics, Tsinghua University, Beijing 100084, China}

\begin{abstract}
As an extended and more complete version of the primary QCD nucleation model presented
in Ref. \cite{Feng:2016ddr}, the new model introduces explicitly the transition balance and 
formulates it in both macroscopic and microscopic descriptions. The microscopic description
of the transition balance in QCD nucleation is implemented in a kinetic parton cascade model
and tested for a first-order phase transition from gluons to pions in a one-dimensional
expansion with Bjorken boost invariance.
\end{abstract}

\maketitle

\section{Introduction}
\label{sec1:intro}
One of the common scientific purposes of relativistic heavy-ion collision experiments
at Relativistic Heavy Ion Collider (RHIC) \cite{Odyniec:2013aaa}, Facility for Antiproton and Ion Research (FAIR) \cite{STAR_NOTE_26,Senger:2022bjo}, and 
Nuclotron-base Ion Collider fAcility (NICA) \cite{Kekelidze:2016hhw,Burmasov:2022clm}  is to find the QCD critical point,
which is the end point of the first-order phase transition line in the temperature-baryon
chemical potential plane \cite{Stephanov:1998dy,Stephanov:1999zu,Fodor:2004nz,Allton:2005gk,Gavai:2008zr,deForcrand:2008zi,Hatsuda:2006ps,Braun-Munzinger:2015hba,Braun-Munzinger:2008szb,Fukushima:2013rx}. At this point the largest fluctuation will occur, which
will become a signal of the existence of the critical point in experiments.
Abundant investigations have been made from both theoretical \cite{Stephanov:2011pb,Athanasiou:2010kw,Schmidt:2010xm,Cheng:2008zh,Friman:2011pf,Borsanyi:2018grb,Stephanov:2011pb,Parotto:2018pwx,Stephanov:2017ghc,Rajagopal:2019xwg,Asakawa:2009aj,Mukherjee:2015swa,Gazdzicki:2015ska,Mukherjee:2016kyu} and experimental side \cite{ NA49:2007weq,STAR:2013gus, Mackowiak-Pawlowska:2014ipa,NA61SHINE:2015uhh, PHENIX:2015tkx,Luo:2015ewa,Luo:2017faz,Esha:2017dce,STAR:2017tfy,Prokhorova:2018tcl,Xu:2018vnf,Andronov:2018bln,Behera:2018wqk,STAR:2020tga,STAR:2020ddh,STAR:2021iop,STAR:2021fge,STAR:2021rls,Aparin:2022jok}, see \cite{Stephanov:2006zvm,Bzdak:2019pkr,An:2021wof} for an overview.

While the transition from QGP to hadron gas in an experiment with a certain collider energy
may not meet the critical point, more probably it may cross the first-order phase transition
line. The experimental evidence of the existence of the first-order phase transition is
equivalent to that of the critical point. Therefore, it is important to study the dynamics of
QCD nucleation \cite{Csernai:1992tj,Kapusta:1994pg} and the experimental signature
of the first-order phase transition in relativistic heavy-ion collisions. For this purpose
we plan to build up a hydrodynamic and a kinetic transport model, which can describe
the evolution of the separate phases in first-order phase transition.

The first step has been made in Ref. \cite{Feng:2016ddr}, where we have established 
a framework of hydrodynamic description of separated phases in the first-order
phase transition from gluons to pions. Based on this, we have developed a microscopic
scheme of the transition within the kinetic parton cascade model, Boltzmann Approach of
MultiParton Scatterings (BAMPS) \cite{Xu:2004mz,Xu:2007aa,Uphoff:2014cba}. The dynamical scheme has been proven by comparisons of the numerical results with
the analytical solutions in a one-dimensional expansion of a dissipative fluid with
Bjorken boost invariance.

However, the established framework needs improvements. First, the first-order phase
transition from gluons to pions at the fixed transition temperature is not realistic for
the QGP created in relativistic heavy-ion collisions. The transition from all parton species
to all hadron species at finite baryon chemical potential should be considered. Moreover,
one has to take into account that during the first-order transition, the temperature and baryon
chemical potential change accordingly along the first-order phase transition 
line \cite{Feng:2018anl}. Second, the framework applies only for transitions in an expansion, 
but not for those in a compression, in which transitions from hadrons to partons occur.
Thus, the introduced hydrodynamic description is not complete.

In the present paper we extend our previous work \cite{Feng:2016ddr} to be able to
describe QCD nucleation in both an expansion and a compression. 
We assume, as done in Ref. \cite{Feng:2016ddr}, constant temperature and baryon
chemical potential during the first-order phase transition.
In section \ref{sec2:model} we introduce
our nucleation model. Compared with the primary model in Ref. \cite{Feng:2016ddr}
we take formally the particle or heat flow into account. The hydrodynamic description of
the transition balance is presented in section \ref{sec3:detailedbalance}. The main idea
is that both the transition from partons to hadrons and the inverse transition from hadrons
to partons occur simultaneously. This gives a new picture of QCD 
nucleation. Within this picture, not only the growth of hadron bubbles in an expansion,
but also their shrinkage in a compression can be described. Based on the hydrodynamic
description, in section \ref{sec4:micro} we establish a new microscopic
scheme of QCD nucleation by using the similar procedure as given in 
Ref. \cite{Feng:2016ddr}. The microscopic scheme will be implemented in BAMPS and
tested in section \ref{sec5:test} for the first-order phase transition from gluons to pions 
in a longitudinal expansion with Bjorken boost invariance. The numerical results are shown
and compared with analytical ones derived in Ref. \cite{Feng:2016ddr}. Finally, we give
a summary and an outlook in section \ref{sec6:summary}.

\section{Hydrodynamic description of QCD nucleation}
\label{sec2:model}
Since the degrees of freedom of partons are more than those of hadrons, the first-order
phase transition from the partonic phase to hadronic phase can only happen in such spatial
region, where the matter expands. Vice versa, the first-order phase transition from
the hadronic phase to partonic phase can only happen in compressing regions.
Only in this way the entropy increases or maintains, so that the second law of thermodynamics
holds.

A QCD nucleation will occur, when the temperature reaches the transition temperature $T_c$.
Like the boiling water, bubbles containing hadrons will appear.
If the temperature fluctuates spatially, we can imagine the occurrence of a number
of hadron bubbles with different sizes. Each hadron bubble is separated from the QGP by
an interface, which is a mixture of partons and hadrons. Transitions from partons to
hadrons will enlarge the hadron bubble, while transitions from hadrons to partons will shrink
the hadron bubble. In both cases, the physical process of the phase transition occurs in the
interface, where the Gibbs condition holds.

\begin{figure}[t]
	\centering
	\includegraphics[width=0.4\textwidth]{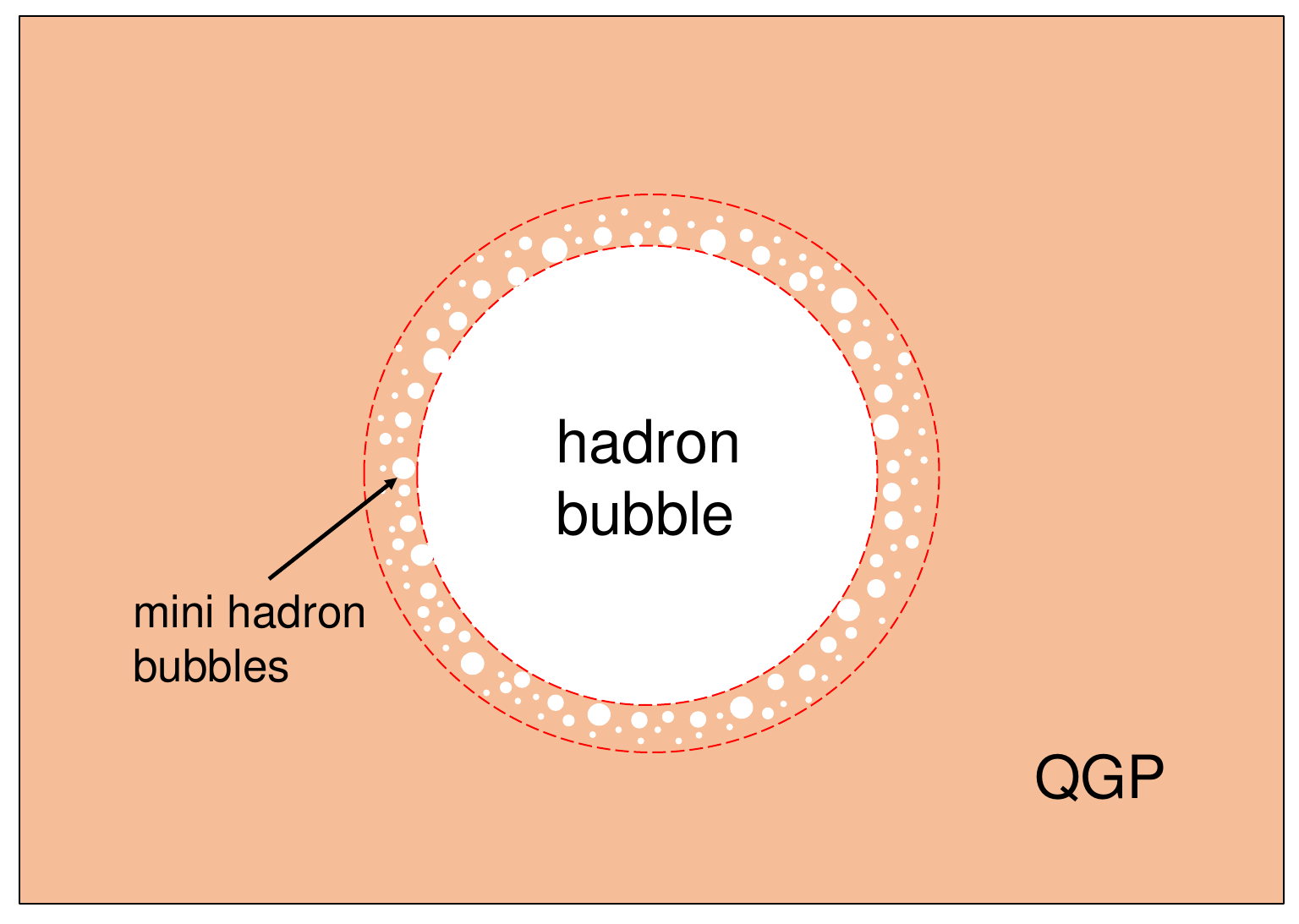}
	\caption{(Color online) The illustration of a hadron bubble with its interface.}
	\label{illustration}
\end{figure}

How the phase transition in the interface proceeds, is not clear yet, the same as
the physical understanding for the creation of hadron bubbles \cite{Csernai:1992tj}.
Our idea is the following.
We assume that there exists mini hadron bubbles in the interface, as illustrated
in Fig. \ref{illustration}. 
Although the mini hadron bubbles are much smaller than hadron bubbles, they can 
still be regarded as macroscopic objects. The mini hadron bubbles may evolve with
increasing sizes and merge to the hadron bubble, which makes it bigger.
Also, they may evolve with decreasing sizes and disappear, which makes the hadron
bubble smaller. Different from hadron bubbles, which only appear during the phase transition,
mini hadron bubbles may have certain probability to appear in QGP even at higher
temperatures than $T_c$. Once a mini hadron bubble appears at $T > T_c$,
it will be squeezed, until it disappears, since the pressure inside the mini hadron
bubble is smaller than that outside. The small size of the mini hadron bubble 
makes the period from the appearance to its disappearance so short that the mini
hadron bubble may not be visible in macroscopic view. When the temperature decreases to
$T_c$, mini hadron bubbles can eventually get larger to form visible hadron bubbles.
Similarly, mini parton bubbles may also appear in hadron gas
at lower temperatures than $T_c$. 

In the rest of this section we give the hydrodynamic description of QCD nucleation by
assuming that the evolution of mini hadron bubbles is governed by hydrodynamic equations. 
As an improvement to our primary work in Ref. \cite{Feng:2016ddr} we take
the particle flow (or heat flow) into account.
 
We now remind the basic hydrodynamic equations. The most general
decomposition of the particle four-flow and the energy-momentum tensor can be written as \cite{DeGroot:1980dk}
\begin{eqnarray}
\label{nmu}
N^\mu&=&n U^\mu + Q^\mu \,, \\
\label{tmunu}
T^{\mu\nu}&=&eU^\mu U^\nu -P\Delta^{\mu\nu}+W^\mu U^\nu +W^\nu U^\mu +\pi^{\mu\nu}\,,
\end{eqnarray} 
where $\Delta^{\mu\nu}=g^{\mu\nu}-U^\mu U^\nu$ with $g^{\mu\nu}=(1,-1,-1,-1)$.
$U^\mu$ is the fluid four-velocity, $n=U_\mu N^\mu$ is the particle number density,
and $e=U_\mu T^{\mu\nu} U_\nu$ is the energy density. 
$P=-\frac{1}{3}\Delta_{\mu\nu}T^{\mu\nu}$ denotes the pressure. $\pi^{\mu\nu}$ is
the shear-stress tensor. The bulk pressure is not taken into account.
$Q^\mu$ and $W^\mu$ denote the flow of particle number and
energy-momentum, respectively. $W^\mu$ vanishes in the Landau's definition of
the fluid four-velocity:
\begin{equation}
U^\mu=\frac{T^{\mu\nu}U_\nu}{\sqrt{U_\alpha T^{\alpha\beta} T_{\beta \gamma}U^\gamma}}\,,
\end{equation} 
which is used throughout this paper.
With the energy-momentum conservation $\partial_\mu T^{\mu\nu}=0$ we obtain
\begin{equation}
\label{vhydro_e}
De=-(e+P)\partial_\mu U^\mu +\pi^{\mu\nu} \nabla_{<\mu} U_{\nu>}\,, 
\end{equation}
where
\begin{eqnarray}
&&D=U^\mu \partial_\mu \,, \\
&&\nabla^\mu=\Delta^{\mu\nu} \partial_\nu \,, \\
&&A^{<\mu\nu>}=\left [ \frac{1}{2} \left ( \Delta^\mu_\alpha \Delta^\nu_\beta
+\Delta^\nu_\alpha \Delta^\mu_\beta \right ) -\frac{1}{3} \Delta^{\mu\nu} \Delta_{\alpha\beta}
\right ] A^{\alpha\beta} \,.
\end{eqnarray}
With the definition of the shear pressure
\begin{equation}
\label{shearpressure}
\tilde \pi=- \frac{\pi^{\mu\nu} \nabla_{<\mu} U_{\nu>}}{\partial_\mu U^\mu}
\end{equation}
Eq. (\ref{vhydro_e}) is reduced to
\begin{equation}
\label{vhydro_e2}
De=-(e+P+\tilde\pi)\partial_\mu U^\mu \,. 
\end{equation}
For the number density we have 
\begin{equation}
\partial_\mu N^\mu =J \,,
\end{equation}
where $J$ is the source of the particle production. Putting Eq. (\ref{nmu})
in the above equation, we obtain
\begin{equation}
\label{vhydro_n}
Dn=-n\partial_\mu U^\mu-\partial_\mu Q^\mu +J \,.
\end{equation}

Considering the occurrence of the first-order phase transition in a piece of a fluid 
(in the interface of a hadron bubble) with a volume of $V$ in its rest frame 
at the proper time $\tau$. The total volume of mini hadron
bubbles is $V_h$, while the volume of partons is $V_p$. We have 
$V=V_p+V_h$ and $f_p=V_p/V$ is defined as the partonic fraction.
Thus, the total particle number and energy density read,
\begin{eqnarray}
n&=&f_p n_p +(1-f_p) n_h \,, \\
e&=&f_p e_p +(1-f_p) e_h \,,
\end{eqnarray}
where $n_p$ and $n_h$ ($e_p$ and $e_h$) denote the number (energy) density of
partons and hadrons, respectively. 
Furthermore, $Q^\mu$ and $\tilde \pi$ can be separated into the partonic and hadronic part,
\begin{eqnarray}
Q^\mu&=&f_p Q^\mu_p +(1-f_p) Q^\mu_h \,, \\
\tilde \pi&=&f_p \tilde \pi_p +(1-f_p) \tilde \pi_h \,.
\end{eqnarray}
After a time $d\tau$, the volumes are changed to $V+dV$,
$V_p+dV_p$, and $V_h+dV_h$ with $dV=dV_p+dV_h$.
We assume that during the phase transition,  both the parton and hadron phase
have the same fluid four-velocity $U^\mu$. The time evolution of total 
particle number and energy density are governed by Eqs. (\ref{vhydro_n}) and
(\ref{vhydro_e2}).
Taking the derivative $Dn$ in Eq. (\ref{vhydro_n}) in the local rest frame, we have
\begin{equation}
Dn=\frac{d n}{d \tau}=\frac{d}{d \tau} \frac{N}{V}
=-n\frac{1}{V}\frac{d V}{d\tau} +\frac{1}{V} \frac{d N}{d \tau} \,.
\end{equation}
Thus,
\begin{eqnarray}
\label{dV}
&&\frac{1}{V} \frac{dV}{d\tau}=\partial_\mu U^\mu \,, \\
&&\frac{1}{V} \frac{d N}{d \tau}=-\partial_\mu Q^\mu + J\,.
\end{eqnarray}
$J$ corresponds to the number changing in the parton and hadron phase as well as
in the conversion between partons and hadrons, while $-\partial_\mu Q^\mu$ indicates
the particle flow. We denote $dN^F$ as the number of the flowing partons and hadrons
during $d\tau$, which gives
\begin{equation}
\label{dN}
\frac{1}{V} \frac{d N^F}{d \tau}=-\partial_\mu Q^\mu \,.
\end{equation}
The volumes occupied by the flowing partons and hadrons are respectively
\begin{eqnarray}
\label{dvpf}
dV_p^F&=&f_p dV^F=f_p \frac{dN^F}{n}\,, \\
\label{dvhf}
dV_h^F&=&(1-f_p) dV^F=(1-f_p) \frac{dN^F}{n} \,.
\end{eqnarray}
Correspondingly, the volume changes due to the phase transition
are denoted by $dV_p^T$ and $dV_h^T$, respectively. We have then 
$dV_p=dV_p^F+dV_p^T$ and $dV_h=dV_h^F+dV_h^T$.

Taking the derivative $De$ in Eq. (\ref{vhydro_e2}) in the local rest frame and
using Eq. (\ref{dV}) we obtain $dE= -(P+ \tilde \pi) dV$, which indicates that
the total energy in $V$ changes during $d\tau$ due to the work done by the pressure.
$dE$ is negative in an expansion ($dV > 0$), while it is positive in a compression 
($dV < 0$). We have required here that $P+\tilde \pi >0$. 
According to the Gibbs condition, the parton and hadron energy densities
are constant during the phase transition. Thus, we have
$dE=e_p dV_p+e_h dV_h$, then obtain the energy balance
\begin{equation}
\label{energy}
-e_p dV_p = e_h dV_h + (P+ \tilde \pi) dV \,.
\end{equation}
It shows that the energy lost in the parton phase is transferred in the hadron phase and 
compensates the energy loss due to the work done by the pressure in an expansion
($dV > 0$) for instance. With $dV_h=dV-dV_p$ we obtain
\begin{equation}
\label{dvp}
dV_p=dV_p^T+dV_p^F=-\frac{e_h+P+\tilde \pi}{e_p-e_h} dV \,.
\end{equation}
This equation determines the change of the spatial occupation of the partonic 
phase (and of the mini hadron bubbles $dV_h$) during a time $d\tau$. Equivalently,
from the definition of $f_p$, we have
\begin{equation}
\frac{df_p}{d\tau}=\frac{d}{d\tau} \frac{V_p}{V}=\left (-\frac{e_h+P+\tilde \pi}{e_p-e_h} 
-f_p \right ) \partial_\mu U^\mu \,.
\end{equation}
It shows that the evolution of the partonic fraction depends on the rate of expansion
(or compression) $\partial_\mu U^\mu$ and the shear pressure $\tilde \pi$ only,
when the temperature, pressure, and chemical potential are fixed during the phase transition. 

The actual transition between partons and hadrons is determined by $dV_p^T$
and $dV_h^T$. From Eqs. (\ref{dN}), (\ref{dvpf}), (\ref{dvhf}), and (\ref{dvp}) we obtain
\begin{eqnarray}
\label{dvpt}
dV_p^T &=& \left( -\frac{e_h+P+\tilde \pi}{e_p-e_h}
+\frac{f_p}{n} \frac{\partial_\nu Q^\nu}{\partial_\mu U^\mu} \right ) dV \,, \\
\label{dvht}
dV_h^T &=& \left( \frac{e_p+P+\tilde \pi}{e_p-e_h}
+\frac{1-f_p}{n} \frac{\partial_\nu Q^\nu}{\partial_\mu U^\mu} \right ) dV \,.
\end{eqnarray}
In an expansion $(dV > 0)$, where partons are converted into hadrons, $dV_p^T$
should be negative and $dV_h^T$ should be positive. Vice versa, in a compression 
$(dV < 0)$, where
hadrons are converted into partons, $dV_p^T$ should be positive and $dV_h^T$ should be
negative. This provides the conditions for the applicability of our QCD
nucleation model,
\begin{eqnarray}
\label{condition1}
\mbox{Condition:}\ \ &&\nonumber \\
&&  -\frac{e_h+P+\tilde \pi}{e_p-e_h}
+\frac{f_p}{n} \frac{\partial_\nu Q^\nu}{\partial_\mu U^\mu} < 0 \\
\label{condition2}
\mbox{and } && \frac{e_p+P+\tilde \pi}{e_p-e_h}
+\frac{1-f_p}{n} \frac{\partial_\nu Q^\nu}{\partial_\mu U^\mu} >0  \,.
\end{eqnarray}
These conditions are obviously fulfilled in ideal fluids, since $\tilde \pi$ and 
$\partial_\nu Q^\nu$ vanish. Therefore, these conditions
shall be fulfilled, if the particle system is not too far from thermal equilibrium.

\section{Transition balance in QCD nucleation}
\label{sec3:detailedbalance}
Suppose a QCD nucleation via the first-order phase transition in an expanding medium.
In the previous section we have derived the volume changes of the parton and hadron phase
for the conversion, i.e., $dV_p^T$ and $dV_h^T$ from Eqs. (\ref{dvpt}) and (\ref{dvht}).
In the considered case, $dV_p^T$ is negative, while $dV_h^T$ is positive. Partons are
converted into hadrons. In the expansion, both the parton
and hadron phase should gain energy to balance the energy loss due to the work outwards, 
in order to maintain the temperature and pressure in both phases. It is true
for the hadron phase, because accompanying the conversion from partons to hadrons,
energy will enter the hadron phase and can eventually balance the
energy loss of the hadron phase. On the other hand, it is not obvious for the parton phase,
how to gain enough energy to balance its energy loss. 
In our primary model \cite{Feng:2016ddr}, we have implemented microscopic back
reactions (from hadrons to partons) by hand to balance the energy loss in the parton phase.
Thus, the primary model is established only for
the transition from partons to hadrons in an expansion, but not for the one from hadrons
to partons in a compression.

In this section we extend our primary model to a complete hydrodynamic form, which 
can describe both the transitions from partons to hadrons as well as from hadrons to 
partons. To this end we propose simultaneous transitions from partons to hadrons and
from hadrons to partons. We imagine that during a time interval some mini hadron bubbles expand and some other mini hadron bubbles shrink. If the whole system is expanding,
the increase of the mini hadron bubbles is dominant, while if the whole system is shrinking, 
the decrease of the mini hadron bubbles is dominant. The net change of all mini hadron
bubbles is governed by the description in the previous section,
see Eqs. (\ref{dvpt}) and (\ref{dvht}). The shrinking mini hadron bubbles guide
the hydrodynamic description of the transition from hadrons to partons, which is lacking
in the primary model \cite{Feng:2016ddr}.

We denote $dV_p^{PtH}$ and $dV_h^{PtH}$ as the volume changes of
the parton and hadron phase in the transition from partons to hadrons (PtH stands for the
conversion from partons to hadrons), while we denote $dV_p^{HtP}$ and $dV_h^{HtP}$ as
those of the transition from hadrons to partons.
Thus, per definition, $dV_p^{PtH}$ and $dV_h^{HtP}$ are always negative, while
$dV_h^{PtH}$ and $dV_p^{HtP}$ are always positive. They are related by
\begin{equation}
dV_p^T=dV_p^{PtH}+dV_p^{HtP}\,, \ \ 
dV_h^T=dV_h^{PtH}+dV_h^{HtP} \,.
\end{equation}
We define $\alpha \equiv dV_h^T/dV_p^T$, $\alpha_{PtH}\equiv dV_h^{PtH}/dV_p^{PtH}$, 
and $\alpha_{HtP} \equiv dV_h^{HtP}/dV_p^{HtP}$. In the case that there is no net transition, we have $dV_p^T=dV_h^T=0$, which leads to $\alpha_{PtH}=\alpha_{HtP}$.
We assume that $\alpha_{PtH}=\alpha_{HtP}$ should still hold,
when there is a net transition between partons and hadrons. Then it follows that
$\alpha=\alpha_{PtH}=\alpha_{HtP}$. For ideal fluids, where $\tilde \pi$ and 
$\partial_\nu Q^\nu$ vanish, we obtain
\begin{equation}
\label{alpha}
\alpha_{PtH}=\alpha_{HtP}=\alpha = - \frac{e_p +P}{e_h + P} > -\frac{e_p}{e_h}
\end{equation}
according to  Eqs. (\ref{dvpt}) and (\ref{dvht}). Since $dV$ in Eqs. (\ref{dvpt}) and (\ref{dvht}) cancel out, equation and inequality (\ref{alpha}) hold, no matter in an expansion or
a compression. We assume that the inequality 
$\alpha_{PtH}=\alpha_{HtP}=\alpha > -e_p/e_h$ still holds in viscous fluids.

Following the energy balance in Eq. (\ref{energy}), we build up the energy balance in
the transition from partons to hadrons and in the inverse transition, respectively,
\begin{eqnarray}
\label{pth}
-e_p dV_p^{PtH} =&& e_h dV_h^{PtH}+e_h dV_h^F+ (P+ \tilde \pi_h) (1-f_p)dV \nonumber \\
&&+dq \,, \\
\label{htp}
-e_h dV_h^{HtP} = && e_p dV_p^{HtP} +e_p dV_p^F + (P+ \tilde \pi_p) f_p dV \nonumber \\
&& -dq \,.
\end{eqnarray}
The sum of Eqs. (\ref{pth}) and (\ref{htp}) gives Eq. (\ref{energy}), so that the evolution of the
total energy is guaranteed. $dV$ is either positive in an expansion or negative in
a compression. Also, $dV_p^F$ and $dV_h^F$ can be positive or negative for the particle
flow from or into the neighborhood of $V$.

Equation (\ref{pth}) shows the energy transfer from the partonic phase to the expanding mini
hadron bubbles. Because of Eq. (\ref{alpha}), we have
\begin{equation}
\label{ineq1}
-e_p dV_p^{PtH} > e_h dV_h^{PtH} \,.
\end{equation}
The energy of lost partons should be larger than the energy of gained hadrons, no
matter in an expansion or a compression. It can easily be understood in an expansion, since
the energy excess $-e_p dV_p^{PtH} - e_h dV_h^{PtH}$ must at least compensate
the energy loss of the hadron phase due to the work outwards, 
$(P+ \tilde \pi_h) (1-f_p)dV > 0$. In a compression, this term is negative. To hold
the inequality (\ref{ineq1}) mathematically, a positive energy $dq$ must be present 
and sufficient large.

To complete the energy balance, we add the term $e_h dV_h^F$, which can be positive or
negative. A positive $dV_h^F$ indicates that hadrons with number $n_h dV_h^F$ flow
from the neighborhood into the volume $V$. To compensate the accompanying energy win,
an energy of $e_h dV_h^F$ will flow out of $V$, so that there is no net energy flow
in the Landau definition of the flow velocity.  Analogously, for a negative $dV_h^F$,
hadrons with number $-n_h dV_h^F$ flow out of $V$. To compensate the energy loss,
an energy of $-e_h dV_h^F$ from the neighborhood of $V$ will flow into $V$.

Analogous to Eq. (\ref{pth}), Eq. (\ref{htp}) describes the energy transfer from the shrinking 
mini hadron bubbles to the partonic phase. We have
\begin{equation}
\label{ineq2}
-e_h dV_h^{HtP} < e_p dV_p^{HtP} \,.
\end{equation}
Moving $-dq$ to the left hand side of Eq. (\ref{htp}), we see that an energy of $dq$ must be
added into the hadron phase during the conversion, in order to provide enough energy
for the converted partons to compensate the energy loss
due to the particle flow and the work done by the pressure in the parton phase in an expansion.

From Eqs. (\ref{pth}) and (\ref{htp}) with $\alpha_{PtH}=\alpha_{HtP}=\alpha$, we obtain
\begin{eqnarray}
\label{dvpth}
-dV_p^{PtH} =&& \frac{e_h dV_h^F+ (P+ \tilde \pi_h) (1-f_p)dV+dq}
{e_p+\alpha e_h} \,, \\
\label{dvhtp}
-dV_h^{HtP} = && \alpha \frac{e_p dV_p^F + (P+ \tilde \pi_p) f_p dV-dq}
{e_p+\alpha e_h}\,.
\end{eqnarray}
Thus, $dV_p^{PtH}$, $dV_h^{PtH}=\alpha dV_p^{PtH}$, $dV_h^{HtP}$, and 
$dV_p^{HtP}=dV_h^{HtP}/\alpha$ are determined, when $dq$ is fixed.
To hold both the inequalities (\ref{ineq1}) and (\ref{ineq2}),
$dq$ in Eqs. (\ref{pth}) and (\ref{htp}) must be larger than 
$-e_h dV_h^F- (P+ \tilde \pi_h) (1-f_p)dV$ and $e_p dV_p^F + (P+ \tilde \pi_p) f_p dV$.

Some comments on the meaning of the quantity $dq$ are given below. First,
$dq$ is a positive parameter, which is mathematically needed to hold the
inequalities (\ref{ineq1}) and (\ref{ineq2}). With $dq$, in any transitions, no matter
they are from partons to hadrons or from hadrons to partons and they occur in an expansion
or a compression, the energy of partons is always larger than that of hadrons.
Second, from Eqs. (\ref{pth}) and (\ref{htp}) we see that $dq$ is the
energy which is released from those partons during their transition to hadrons,
while it is absorbed by those hadrons, before they are converted to partons. Thus, $dq$
can be understood as a kind of latent heat,
which is exchanged between the parton and hadron phase during the simultaneous
transitions. Different from the latent heat from text books, $dq$ is a quantity in
a microscopic sense. Third, the amount of $dq$
affects how many partons and hadrons are involved in the phase transition. From
Eqs. (\ref{dvpth}) and (\ref{dvhtp}) we see that the larger $dq$, the more partons
are converted into hadrons (larger $-dV_p^{PtH}$) and the more hadrons are converted
into partons (larger $-dV_h^{HtP}$). $dq$ can also be considered as
a phenomenological parameter, which affects the probability that a microscopic
transition process occurs.
Fourth, following the third point, $dq$ affects the fluctuation in particle number
during the phase transition, which is a macroscopic effect. On the other hand, on average,
$dq$ has no macroscopic effect on the phase transition,
since the sum of Eqs. (\ref{pth}) and (\ref{htp}) should be equal to  
Eq. (\ref{energy}) and  $dq$s cancel.

\section{Microscopic scheme of QCD nucleation}
\label{sec4:micro}
Based on Eqs. (\ref{pth}), (\ref{htp}), (\ref{dvpth}), and (\ref{dvhtp}) given in the previous
section, we now establish a new microscopic scheme for QCD nucleation. The primary
scheme has been presented in Ref. \cite{Feng:2016ddr}, where the microscopic
processes from hadrons to partons were implemented by hand.

In microscopic calculations we do not simulate the increase or decrease of mini hadron
bubbles. In the interface partons and hadrons are mixed randomly. The microscopic 
transition processes will give the correct number and kinetic energy of the 
converted particles. 

The space-time evolution of partons and hadrons is described by the kinetic Boltzmann
equations, 
\begin{equation}
p^\mu \partial_\mu f_i (x,p)=C_i[f] \,,
\end{equation}
where $f_i$ is the phase space distribution function of a sort of particles. $i$ stands for 
a parton species or a hadron species. $C_i[f]$ denotes the collision term, which
determines the interactions and transitions involving particles of $i$ at the space-time
point $x\equiv x^\mu$ with the energy-momentum $p\equiv p^\mu$.

We solve the Boltzmann equation by employing the parton transport model Boltzmann
Approach of Multiparton Scatterings (BAMPS) \cite{Xu:2004mz}. 
In BAMPS, particles are sampled in phase space at an initial time. The time evolution
of $f_i$ in phase space is the result of the free streaming of particles and collisions
among particles.  Collisions of particles in a spatial local cell have certain collision
probabilities, which can be derived from the collision term $C_i[f]$. With these collision
probabilities we simulate collisions in a stochastic manner.
In this section, microscopic processes of the conversion from partons
to hadrons and their inverse processes will be constituted, and the probabilities of these processes will be given.

For simplicity, we consider a phase transition from gluons to pions in a one-dimensional
expansion with Bjorken boost invariance \cite{Bjorken:1982qr}, the same as considered
in Ref. \cite{Feng:2016ddr}. The simulation for transitions involving quarks and other hadron
species is more complicated and postponed in a future work.
Furthermore, we assume that there is no particle flow in the considered case, i.e., 
$\partial_\mu Q^\mu_{g/\pi}=0$
and $dV_g^F=dV_\pi^F=0$. From Eqs. (\ref{dvpt}), (\ref{dvht}) we obtain
\begin{equation}
\label{alpha_noflow}
\alpha=\frac{dV_\pi^T}{dV_g^T}=-\frac{e_g+P +\tilde\pi}{e_\pi+P+\tilde\pi} \,.
\end{equation}
Assuming massless pions and Boltzmann statistics for gluons and pions, the energy
densities and pressures are
\begin{eqnarray}
\label{edens}
&& e_g=3n_g T_c +B \,, \ e_\pi=3n_\pi T_c \,, \\
&& P_g=n_g T_c -B \,, \ P_\pi = n_\pi T_c \,. 
\end{eqnarray}
where $B^{1/4}=0.23$ GeV is the bag constant \cite{Chodos:1974je}.
The Gibbs condition gives $P_g=P_\pi$.
The ratio of the number of gained pions to that of lost gluons in the conversion from
gluons to pions is then
\begin{equation}
\label{nratio}
r=\frac{n_\pi dV_\pi^{PtH}}{-n_g dV_g^{PtH}}=-\frac{n_\pi}{n_g}\alpha 
=1+ \frac{(n_\pi-n_g)\tilde\pi}{n_g(e_\pi+P+\tilde\pi)} \,.
\end{equation}
Due to the assumption $\alpha_{PtH}=\alpha_{HtP}=\alpha$, $r$ is the same
as the ratio of the number of lost pions to that of gained gluons in the conversion from pions
to gluons. According to the definition of the shear pressure Eq. (\ref{shearpressure}),
$\tilde \pi$ is negative for an expansion. $P+\tilde \pi$ should be positive for our
hydrodynamic description of QCD nucleation. Therefore, we have $r > 1$,
which indicates that gained (lost) pions are more than lost (gained) gluons in the 
(inverse) conversion. We need processes such as $2g \to 3\pi$, $2g \to 4\pi$, etc, and
their inverse processes. Moreover, due to $P+\tilde \pi>0$ we have 
\begin{equation}
\alpha> -\frac{e_g}{e_\pi}  =-\frac{4n_g-n_\pi}{3n_\pi}\,.
\end{equation}
Thus,
\begin{equation}
\label{ineq3}
1< r=-\frac{n_\pi}{n_g} \alpha < \frac{4n_g-n_\pi}{3n_g}=\frac{61}{48}  \,,
\end{equation}
where  the degeneracy factors of gluons ($d_g=16$) and pions ($d_\pi=3$) are used.  
Due to $r<3/2$, the processes $g+g \leftrightarrow \pi+\pi$ and 
$g+g \leftrightarrow \pi+\pi+\pi$ are sufficient to complete the conversion.


We denote the probability of $g+g\to \pi+\pi$ and 
$g+g\to \pi+\pi+\pi$ by $P_{22}$ and $P_{23}$, and the probability of $\pi+\pi \to g+g$ and
$\pi+\pi+\pi \to g+g$ by $P_{22b}$ and $P_{32}$, respectively. The angular distribution of all
the transition processes is assumed to be isotropic. In the following we will derive
the explicit form of these transition probabilities.

We assume that the transition probabilities are independent of the momenta of particles 
involved in the processes. Suppose $N_g$ is the number of gluons 
and $N_\pi$ is the number of pions in the volume $V$. Thus, the number of
permutations for a two gluons pair is $N_g(N_g-1)/2$. The numbers for a two pions pair
and a three pions triplet are $N_\pi(N_\pi-1)/2$ and $N_\pi (N_\pi-1)(N_\pi-2)/6$.
We have then the numbers of lost gluons and gained pions in the conversion from gluons to
pions, i.e., in the processes $g+g\to \pi+\pi$ and $g+g\to \pi+\pi+\pi$,
\begin{eqnarray}
\label{nglost}
&&\frac{1}{2} N_g (N_g-1) \left (2 P_{22}+ 2 P_{23} \right )=-n_g dV_g^{PtH} \,, \\
\label{npgain}
&&\frac{1}{2} N_g (N_g-1) \left (2 P_{22}+ 3 P_{23} \right )=n_\pi dV_\pi^{PtH} \,.
\end{eqnarray}
Solving the above equations we obtain
\begin{eqnarray}
\label{P22}
P_{22}&=&\frac{-3n_g dV_g^{PtH}-2n_\pi dV_\pi^{PtH}}{N_g (N_g-1)} \nonumber \\
&=&-\frac{(3-2r)n_gdV_g^{PtH} }{N_g (N_g-1)} \nonumber  \\
&\approx& -\frac{3-2r}{f_g^2n_g} \frac{dV_g^{PtH}}{V^2} \,, \\
\label{P23}
P_{23}&=&2\frac{n_g dV_g^{PtH}+n_\pi dV_\pi^{PtH}}{N_g (N_g-1)} \nonumber \\
&=&\frac{2(1-r)n_gdV_g^{PtH} }{N_g (N_g-1)}
=\frac{2(r-1)}{3-2r}P_{22} \,.
\end{eqnarray}
Analogously, the equations for
the conversion from pions to gluons ($\pi+\pi \to g+g$ and $\pi+\pi+\pi \to g+g$ processes)
read
\begin{eqnarray}
\label{nplost}
&&\frac{1}{2} N_\pi (N_\pi-1) 2 P_{22b}+ \frac{1}{6} N_\pi (N_\pi-1)(N_\pi-2) 3 P_{32} \nonumber \\
&=& -n_\pi dV_\pi^{HtP} \,, \\
\label{nggain}
&&\frac{1}{2} N_\pi (N_\pi-1) 2 P_{22b}+ \frac{1}{6} N_\pi (N_\pi-1)(N_\pi-2) 2 P_{32} \nonumber \\
&=& n_g dV_g^{HtP} \,.
\end{eqnarray}
The solutions are
\begin{eqnarray}
\label{P22b}
P_{22b}&=&\frac{3n_g dV_g^{HtP}+2n_\pi dV_\pi^{HtP}}{N_\pi (N_\pi-1)}  \nonumber \\
&=&-\frac{(3/r-2)n_\pi}{N_\pi (N_\pi-1)}dV_\pi^{HtP} \nonumber \\
&\approx& -\frac{(3/r-2)}{(1-f_g)^2n_\pi} \frac{dV_\pi^{HtP}}{V^2} \,, \\
\label{P32}
P_{32}&=&6\frac{-n_g dV_g^{HtP}-n_\pi dV_\pi^{HtP}}{N_\pi (N_\pi-1) (N_\pi-2)} \nonumber \\
&=&-6\frac{(1-1/r)n_\pi}{N_\pi (N_\pi-1)(N_\pi-2)}dV_\pi^{HtP} \nonumber \\
&=&\frac{6(1-1/r)}{(3/r-2)(N_\pi-2)}P_{22b} \nonumber \\
&\approx&\frac{6(r-1)}{3-2r}\frac{1}{(1-f_g) n_\pi V}P_{22b} \,.
\end{eqnarray}
All the transition probabilities are fixed, when $dq$ in Eqs. (\ref{dvpth}) and 
(\ref{dvhtp}) is given. Its setup will be shown in the next section.

We now turn to discuss the momentum-energy transfer in processes 
$g+g \to \pi+\pi$, $g+g \to \pi+\pi+\pi$, and their inverse processes. In BAMPS a gluon
takes only its kinetic energy. But in the transition processes, the potential energy
is also involved. According to the energy balance Eq. (\ref{pth}) and the energy density
Eq. (\ref{edens}), the energy of gluons before the conversion is the sum of the 
kinetic energy $-3n_gT_c dV_p^{PtH}$ and the potential energy $-B dV_p^{PtH}$. The
total energy is equal to the sum of the to be released energy $dq$ and 
the kinetic energy of pions after the conversion. We take 
$-(3n_g T_c+B)dV_g^{PtH}-dq$ as the gluon energy in the processes $g+g \to \pi+\pi$
and $g+g\to \pi+\pi+\pi$ with energy-momentum conservation.
The average energy per to be converted gluon is then
\begin{eqnarray}
\epsilon_g&=&\frac{-(3n_g T_c+B)dV_g^{PtH}-dq}{-n_g dV_g^{PtH}} \nonumber \\
&=&3T_c+\frac{B}{n_g}- \frac{dq}{-n_g dV_g^{PtH}} \nonumber \\
&=&3T_c+\frac{B}{n_g} -\frac{dq}{(P+\tilde \pi_h)(1-f_g)dV+dq}  \frac{e_g+\alpha e_\pi}{n_g} 
\nonumber \\
&=&3T_c+\frac{B}{n_g} -\frac{dq}{(P+\tilde \pi_h)(1-f_g)dV+dq} \nonumber \\
&& \times \left [ \frac{B}{n_g} - (r-1)3T_c \right ] \,. 
\end{eqnarray} 
It is obvious that $\epsilon_g > 3T_c$. Since transition processes with momentum and
kinetic energy conservation are numerically implemented in a standard routine, we {\it amplify}
the momentum (also the kinetic energy) of each gluon by $\epsilon_g/(3T_c)$ before 
performing $g+g\to \pi+\pi$ and $g+g\to \pi+\pi+\pi$ processes. 

For the conversion from pions to gluons we rewrite Eq. (\ref{htp}) as
\begin{eqnarray}
-e_\pi dV_\pi^{HtP} &=& (3n_gT_c+B) dV_g^{HtP} + (n_gT_c-B+ \tilde \pi_g) f_g dV 
\nonumber\\
&& -dq \nonumber \\
&=&3n_gT_c dV_g^{HtP} + (n_gT_c+ \tilde \pi_g) f_g dV \nonumber \\
&&+B(dV_g^{HtP}-f_g dV)-dq \,.
\end{eqnarray}
The potential energy $B(-dV_g^{HtP}+f_g dV)$ and the to be absorbed energy $dq$ 
should be added into the kinetic energy of pions in the conversion. Thus, we take 
$-e_\pi dV_\pi^{HtP}+B(-dV_g^{HtP}+f_g dV)+dq$ as the energy of pions in
the processes $\pi+\pi \to g+g$ and $\pi+\pi+\pi \to g+g$. The average energy per to be
converted pion is
\begin{eqnarray}
\epsilon_\pi &=& \frac{-e_\pi dV_\pi^{HtP}+B(-dV_g^{HtP}+f_g dV)+dq}{-n_\pi dV_\pi^{HtP}} 
\nonumber \\
&=&3T_c+\frac{B}{\alpha n_\pi}-\frac{dq+Bf_g dV}{dq-(P+\tilde \pi_g)f_g dV}
\frac{e_g+\alpha e_\pi}{\alpha n_\pi} \nonumber \\
&=&3T_c-\frac{B}{r n_g}+\frac{dq+Bf_g dV}{dq-(P+\tilde \pi_g)f_g dV} \nonumber \\
&& \times \left [  \frac{B}{r n_g} -\frac{r-1}{r} 3T_c  \right ] \,.
\end{eqnarray}
Because $B=(n_g-n_\pi)T_c$ and $1<r<(4n_g-n_\pi)/(3n_g)$ [see Eq. (\ref{ineq3})], 
$\epsilon_\pi$ is larger than $3T_c$, when $r$ approaches to $1$, while $\epsilon_\pi$ 
becomes smaller than $3T_c$, when $r$ approaches to $(4n_g-n_\pi)/(3n_g)$.
In the latter case, $\pi+\pi+\pi \to g+g$ processes have more contributions than
in the former case. We multiply the momentum (also the kinetic energy) of each pion 
by $\epsilon_\pi/(3T_c)$ before performing $\pi+\pi \to g+g$ and $\pi+\pi+\pi \to g+g$ processes. 

On average, a pion (gluon) after the conversion takes a larger energy than $3T_c$. 
Through collisions with other pions (gluons), these converted pions (gluons) will loose 
the energy excess, which just compensates the energy loss in the pion (gluon) phase due to
the work in the expansion.

\section{Test of the new nucleation model}
\label{sec5:test}
To test the new nucleation scheme that we have introduced in the previous sections,
we demonstrate in this section the first-order phase transition from gluons to pions
by employing BAMPS for a one-dimensional expansion with Bjorken boost invariance. 
The formal transition probabilities and the details of the numerical implementations
of the processes $g+g\to \pi+\pi$, $g+g\to \pi+\pi+\pi$, and their inverse processes are given
in Sec. \ref{sec4:micro}. Except for these, all other setups are almost
the same as those made in Ref. \cite{Feng:2016ddr}. Below we repeat some important setups
and give the heat $dq$ explicitly.

Since the test focuses on the nucleation model during the phase transition, the interactions of
gluons (pions) in the gluon (pion) phase are considered to be elastic with the isotropic angular
distribution for simplicity. The cross section of gluonic scatterings is set to be
$\sigma_g=40 \mbox{ mb}$ and the test particle number per a real particle is set to be
$N_{test}=20000$. Both are larger than those given in Ref.  \cite{Feng:2016ddr}.
At the initial time $\tau_0=0.5 \mbox{ fm/c}$ gluons are distributed according to the Boltzmann 
distribution with the temperature $T_g=0.3 \mbox{ GeV}$. According to Eqs. (52), (58)-(60)
in Ref. \cite{Feng:2016ddr}, the gluon system reaches the temperature at the
phase transition $T_c=0.2227 \mbox{ GeV}$ at $\tau_c=1.415 \mbox{ fm/c}$.
The nucleation begins. At $\tau_c$ the shear viscosity to the entropy density ratio is 
$\eta_g/s_g=0.0406$. The ratio $\eta_g/s_g$ should be maintained during the phase 
transition. We set the cross section of pionic scatterings
$\sigma_\pi=\sigma_g s_g/s_\pi=\sigma_g d_g/d_\pi$,
so that the shear viscosity to the entropy density ratio of pions takes the same value as that
of gluons \cite{Luzum:2008cw,Dusling:2009df,Song:2010aq,Schenke:2010rr,Niemi:2011ix}. With this setup we can obtain analytical $f_g(\tau)$ (see Eq. (45) in 
Ref. \cite{Feng:2016ddr}), with which the numerical result will be compared.

We note that the shear pressure $\tilde \pi$ is not extracted from the particle distribution, but 
is obtained by employing the first-order theory of hydrodynamics. Then the shear tensor reads
\begin{equation}
\pi^{\mu\nu}_{g/\pi}=2\eta_{g/\pi} \nabla^{<\mu} U^{\nu>} \,.
\end{equation}
For the Bjorken expansion, we have
\begin{eqnarray}
\label{hydroexp2}
&& \frac{1}{V} \frac{dV}{d\tau} =\nabla_\mu U^\mu =\frac{1}{\tau} \,, \\
\label{shearpressure2}
&& \tilde \pi_{g/\pi}=-2\eta_{g/\pi} \frac{\nabla^{<\mu} U^{\nu>} \nabla_{<\mu} U_{\nu>}}{\nabla_\mu U^\mu}
=-\frac{4\eta_{g/\pi}}{3\tau} \,.
\end{eqnarray}
According to Eqs. (49), (56), and (57) in Ref. \cite{Feng:2016ddr}, we obtain 
$\eta_g=0.339 \mbox{ fm}^{-3}$ and $P=0.084 \mbox{ GeV fm}^{-3}$ at $\tau_c$. 
The shear pressure is then $\tilde \pi_g =-0.0629 \mbox{ GeV fm}^{-3}$. 
Thus, $P+\tilde\pi_g >0$ and condition (\ref{condition1}) is fulfilled. This is the reason that we
had to use an unrealistic large cross section $\sigma_g$ to get a small $\eta_g$, which leads to an unrealistic small $\eta_g/s_g$ at $\tau_c$. In reality, the phase transition may begin
at a much larger time than that given here. Smaller cross sections
(larger $\eta_g$), with which $P+\tilde \pi_g >0$ [see Eq. (\ref{shearpressure2})], can
apply to our nucleation model.

With $\tilde \pi_g$ and $P$ we calculate $r$ in Eq. (\ref{nratio})
at $\tau_c$ and get $r=1.1872 < 61/48$, which meets the condition (\ref{ineq3}).
Since the absolute value of $\tilde \pi$ is inversely proportional to
$\tau$, $r$ becomes smaller for $\tau > \tau_c$ and condition (\ref{condition1}) and
the inequality (\ref{ineq3}) always hold. This ensures the applicability of our nucleation model.

Remind that $dq$ must be larger than $-e_\pi dV_\pi^F- (P+ \tilde \pi_\pi) (1-f_g)dV$
and $e_g dV_g^F + (P+ \tilde \pi_g) f_g dV$. For our case, $dV_g^F=dV_\pi^F=0$ and
$dV >0$, so that $dq$ must be larger than $(P+ \tilde \pi_g) f_g dV$. We set
$dq=(1+\beta)(P+\tilde \pi_g)f_g dV$ with $\beta=0.2$. Then $dV_g^{PtH}$ and 
$dV_\pi^{HtP}$ in Eqs. (\ref{dvpth}) and (\ref{dvhtp}) are known for given timestep $d\tau$.
With these all the transition probabilities $P_{22}$, $P_{23}$, $P_{22b}$, and $P_{32}$
are fixed.
We note that with larger $\beta$ (also larger $dq$) more particles will 
be involved in the transitions [see Eqs. (\ref{dvpth}) and (\ref{dvhtp})]. This will 
reduce the fluctuation in particle number.

In the following we focus on the local region at zero space-time rapidity with a small
interval of $0.03$ and show the nucleation in this region.
We have checked all the numerical results with the new nucleation model and found that 
they agree with the analytical results, such like those demonstrated 
in Ref. \cite{Feng:2016ddr}. 
Below we will show some of these results, the time evolution of the gluon fraction,
the number and the energy density, and the total entropy.
Other results such as the time evolution of the pressure, temperature, chemical potential,
and entropy density can be obtained from the number and energy densities and will not
be shown in this presentation.

\begin{figure}[t]
 \centering
 \includegraphics[width=0.46\textwidth]{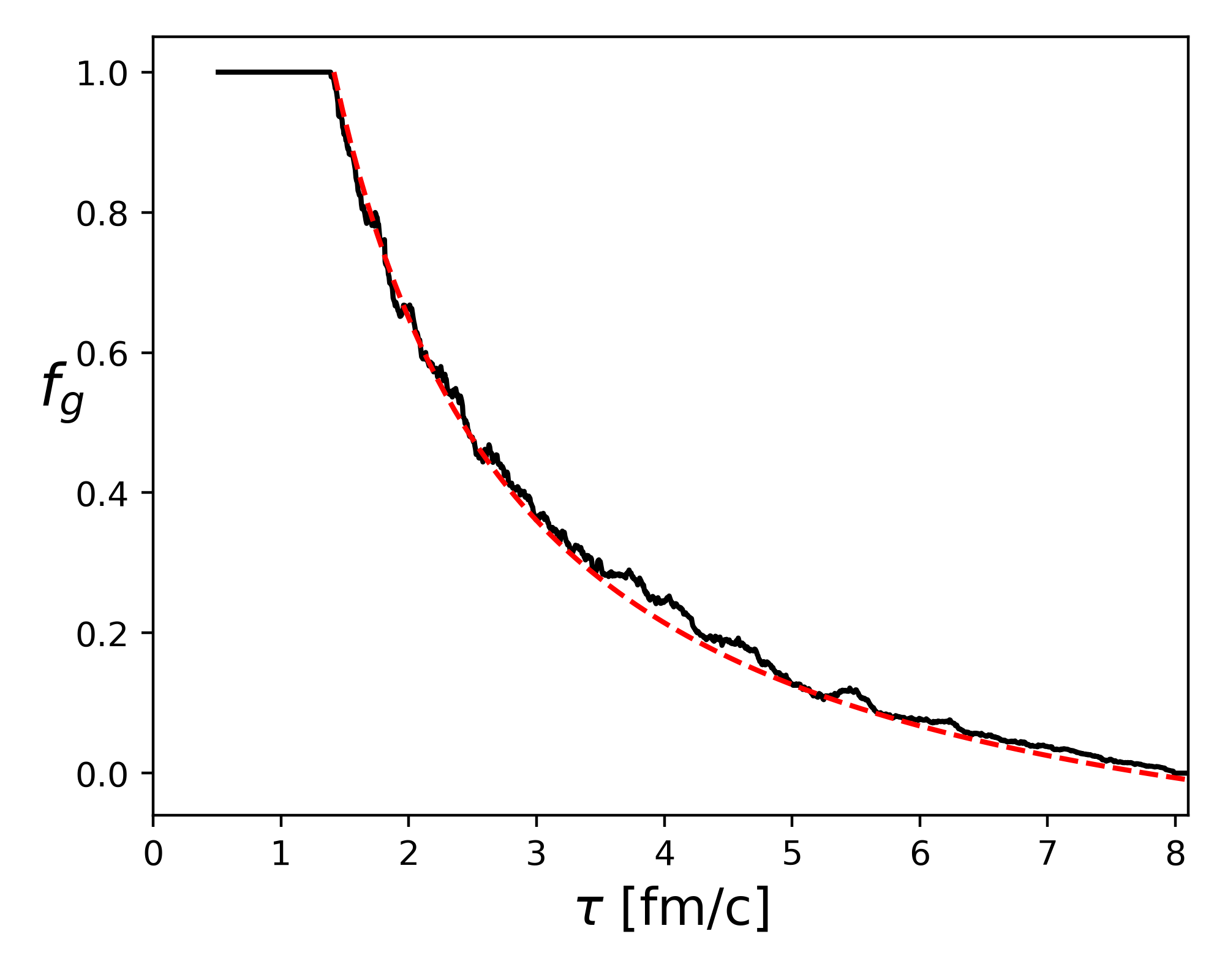}
 \caption{(Color online) The time evolution of the gluon fraction. 
 The solid curve and the dashed curve depict the numerical result and the expected
function respectively.}
 \label{fig2-fg}
\end{figure}

Figure \ref{fig2-fg} depicts the numerical extracted gluon fraction $f_g$ (solid curve)
and the analytical result (dashed curve) according to Eq. (45) in Ref. \cite{Feng:2016ddr}. 
We see a perfect agreement.

\begin{figure}[t]
 \centering
 \includegraphics[width=0.46\textwidth]{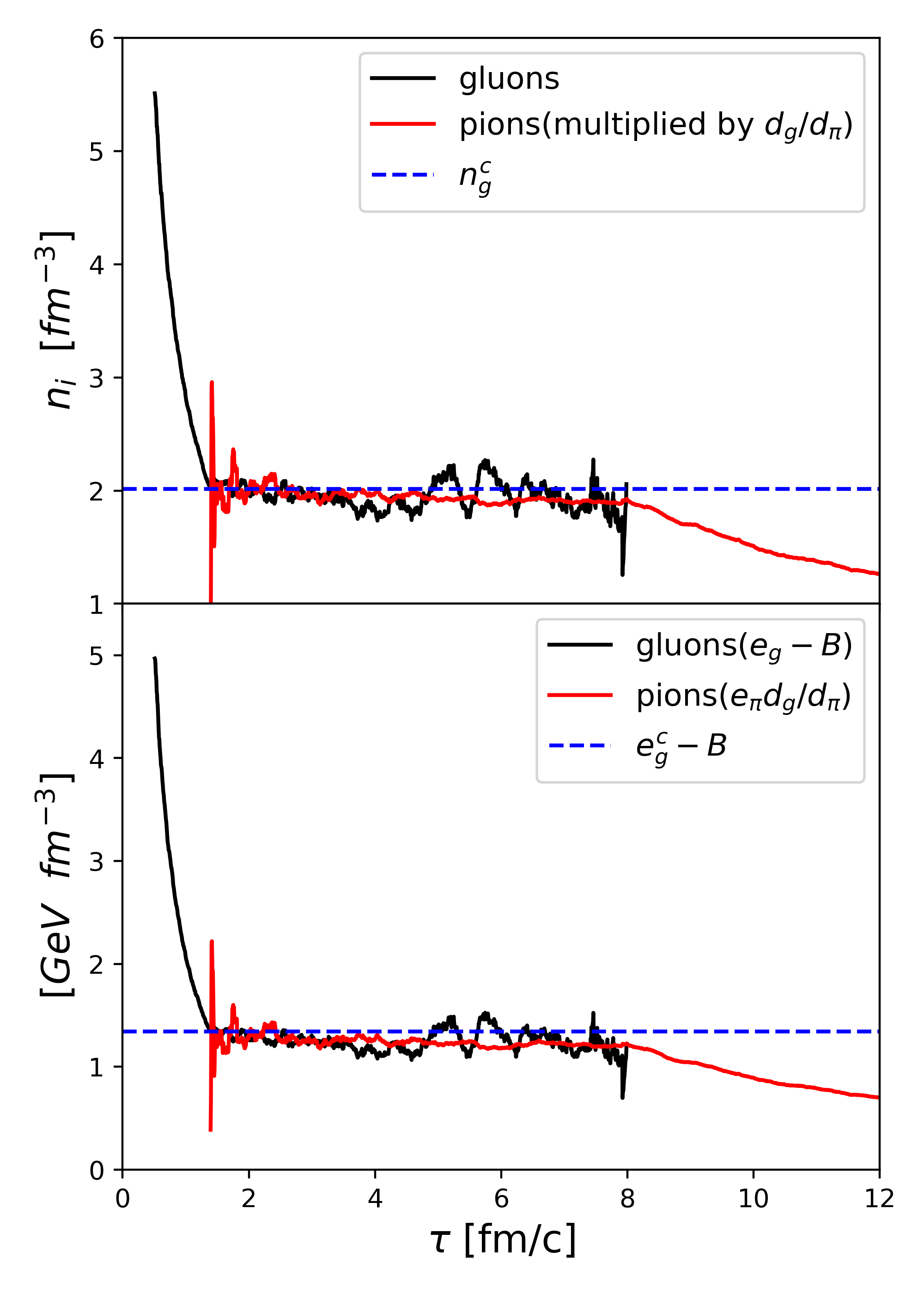}
 \caption{(Color online) The time evolution of the number
and the kinetic energy density. The black (red) curves are for
gluons (pions). The dashed lines depict the analytical values during the phase transition.}
 \label{fig4-twophase1}
\end{figure}

In Fig. \ref{fig4-twophase1} we show the  time evolution
of the number and the kinetic energy density of gluons (black curves) and pions (red curves),
respectively, evaluated by Eqs. (70), (71), and (74) in Ref. \cite{Feng:2016ddr}.
For comparisons, we multiply the densities of pions  by the ratio of the degeneracy factors
$d_g/d_\pi$. We see that the densities keep almost constant during  the phase transition.
However, a slight decrease of the densities is also observed, when comparing
with the analytical values $n_g^c=n_g(\tau_c)=2.013 \mbox{ fm}^{-3}$ and 
$e_g^c-B=e_g(\tau_c)-B=1.3448 \mbox{ GeV fm}^{-3}$, which are denoted by the
dashed lines. This is due to the presence of the particle flow of produced pions to higher rapidities, which is ignored in our present microscopic scheme. The particle flow
lowers the densities and temperature of pions. Subsequently, the conversion from pions
to gluons makes the densities and
temperature of gluons smaller than the analytical values. The effect of particle flow was not
visible in the former scheme \cite{Feng:2016ddr}, because the back reactions to gluons are made by hand.

\begin{figure}[b]
 \centering
 \includegraphics[width=0.46\textwidth]{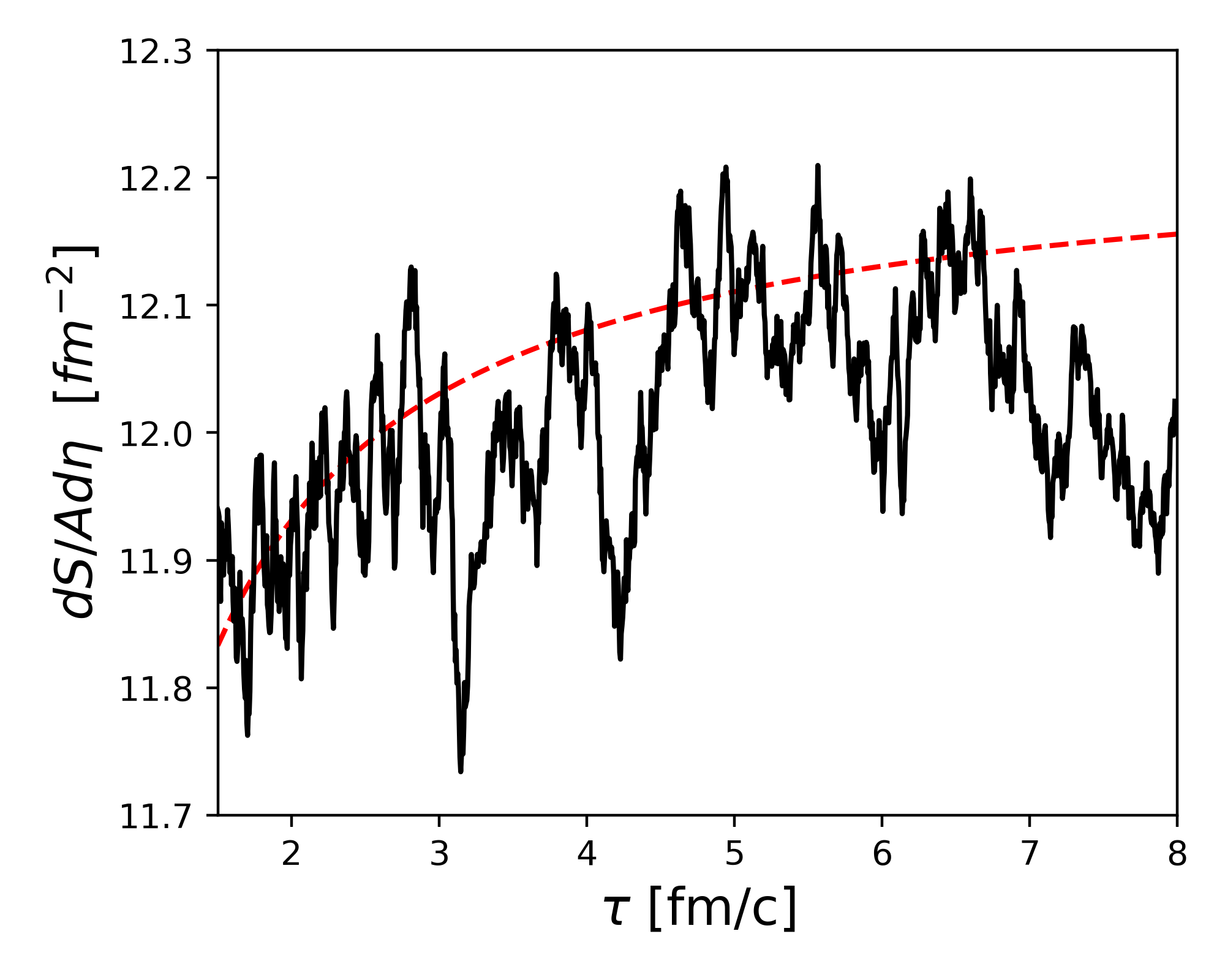}
 \caption{(Color online) The total entropy per space-time rapidity per transverse area.}
 \label{fig7-entropy}
\end{figure}

Finally, the total entropy per space-time rapidity and per transverse area 
is calculated according to Eqs. (24) and (44) in Ref. \cite{Feng:2016ddr} and
is shown in Fig. \ref{fig7-entropy}.
Due to the nonzero but small shear viscosity the increase of 
the total entropy during  the phase transition is not significant, but still visible despite large
numerical uncertainties.

\section{Summary and outlook}
\label{sec6:summary}
In this article we have improved our primary model of QCD nucleation in 
Ref. \cite{Feng:2016ddr} by realizing the transition balance in a formal hydrodynamic
description. The main idea is that the QCD phase transition from one phase to another 
is a result of two reversed and simultaneous transitions:
one is  from partons to hadrons and another one is from hadrons to partons. 
We formulated the particle and energy transfer in both conversions. Based on these,
we established a new dynamic scheme for microscopic processes within the kinetic
theory and implemented in the parton cascade model BAMPS. As already demonstrated
in Ref. \cite{Feng:2016ddr}, we also tested the new and improved scheme by comparing
our numerical
results with the analytical ones derived in Ref. \cite{Feng:2016ddr} for a first-order phase
transition from gluons to pions in a one-dimensional expansion with Bjorken boost
invariance. The numerical and analytical results agree well with each other.

The advantage of the new model of QCD nucleation is that it can describe both 
the phase transition from partons to hadrons in an expansion and the one from hadrons
to partons in a compression. The next step of improvement will be including quarks
and more hadron
species into the model and modeling the first-order phase transition with finite baryon
chemical potential \cite{Feng:2018anl}.

\section*{Acknowledgement}
ZX thanks P. Zhuang for helpful discussions.
This work was financially supported by the National Natural Science Foundation of China
under Grants  No. 12035006 and No. 11890712.
C.G. acknowledges support by the Deutsche Forschungsgemeinschaft (DFG) through
the grant CRC-TR 211 ``Strong-interaction matter under extreme conditions.''
The BAMPS simulations were performed at Tsinghua National Laboratory for Information
Science and Technology and on TianHe-1(A) at National Supercomputer Center in Tianjin.

\bibliographystyle{apsrev}

\end{document}